\documentclass[twoside]{article}
\usepackage{xurl}

\usepackage[T1]{fontenc}
\usepackage[utf8]{inputenc}
\usepackage{breakurl}
\usepackage[breaklinks]{hyperref}
\usepackage{apacdoc} 
\usepackage{natbib}
\usepackage{hyperref}
\setcitestyle{authoryear} 
\usepackage[sc]{mathpazo} 
\usepackage[T1]{fontenc} 
\usepackage{microtype} 

\usepackage[english]{babel} 

\usepackage[hmarginratio=1:1,top=32mm,columnsep=20pt]{geometry} 
\usepackage[hang, small,labelfont=bf,up,textfont=it,up]{caption} 
\usepackage{booktabs} 

\usepackage{lettrine} 

\usepackage{enumitem} 
\setlist[itemize]{noitemsep} 

\usepackage{abstract} 
\usepackage{titlesec} 
\renewcommand \thesection{\arabic{section}}
\renewcommand \thesubsection{\arabic{section}.\arabic{subsection}}
\renewcommand \thesubsubsection{\arabic{section}.\arabic{subsection}.\arabic{subsubsection}}

\titleformat{\section}[block]{\large\scshape\centering}{\thesection.}{1em}{} 
\titleformat{\subsection}[block]{\normalfont\bfseries\normalsize}{\thesubsection.}{1em}{} 
\titleformat{\subsubsection}[block]{\normalfont\itshape\normalsize}{\thesubsubsection.}{1em}{} 

\usepackage{fancyhdr} 
\usepackage{titling} 

\usepackage{hyperref} 

\pretitle{\begin{center}\LARGE\bfseries} 
\posttitle{\end{center}} 
\title{All People, One Sky: A Foundation for IAU CPS Community Engagement} 
\author{%
\\
\textsc{John C.~Barentine}\\[1ex] 
\small Dark Sky Consulting, LLC\\
\small 9420 E Golf Links Rd Ste 108 PMB 237, Tucson, AZ 85730-1317 USA\\ 
\small \href{mailto:john@darkskyconsulting.com}{john@darkskyconsulting.com}\\
\\
\textsc{Jessica Heim}\\[1ex] 
\small University of Southern Queensland Centre for Astrophysics \\
\small West Street, Toowoomba Qld 4350, Australia\\
\small \href{mailto:Jessica.Heim@usq.edu.au}{Jessica.Heim@usq.edu.au}\\
}
\date{} 
\renewcommand{\maketitlehookd}{%
\begin{abstract}
\noindent This report first describes the status quo regarding the emerging deployment of very large groups of  low-Earth-orbit satellites in the late 2010s; the concerns raised by the international astronomy community; and steps the community took to address the issue. We then describe the results of a series of four conferences held in 2020-21 that considered the impacts of large satellite constellations as it impacted a number of stakeholders, and how those outcomes resulted in the establishment of both the IAU Centre for the Protection of the Dark and Quiet Sky from Satellite Constellation Interference (IAU CPS) and its Community Engagement (CE) Hub. We finish with a brief description of CE Hub’s initial plans and activities, flowing from the recommendations of those conferences.
\end{abstract}
}

\begin{document}\sloppy

\maketitle


\section{Introduction}
The decade of the 2020s may be remembered by history, in part, as an era of two distinct but directly connected phenomena: a time of transformation in the use of orbital space in the vicinity of the Earth, and the impact that transformation had on the appearance of the night sky. 

The dawn of the decade saw the rapid proliferation of new satellites launched into Earth orbit by private commercial space interests. Groups of satellites, intended in part to operate as large networks to provide wide-area terrestrial broadband Internet connectivity, were launched in configurations that became known in the media as `megaconstellations’.\footnote{The term appears to have been first mentioned in news stories in around the mid-2010s. See, e.g., Foust, J.~(10 October 2016).~``Mega-constellations and mega-debris''. \emph{The Space Review.} \url{https://www.thespacereview.com/article/3078/1}} New network designs required thousands of objects in low-Earth orbit (LEO) to achieve global coverage with high throughput and low latency. Companies and countries have announced plans to launch over 400,000 satellites for this purpose by 2030.\footnote{Based on news reports, U.S. Federal Communications Commission filings and other sources, we count over 422,000 LEO/MEO satellites proposed for deployment since 2019. Of these, filings for launch and operations clearance exist for about 67,000 satellites. SpaceX and OneWeb alone account for the launch of about 3,700 satellites since 2019.}  As near-Earth space, and especially LEO, quickly fills with these objects, the risks associated with a number of their externalities, from debris-producing collisions to uncontrolled re-entries to unintended geoengineering of the upper atmosphere, are steadily increasing.~\citep{BoleyByers2021,ThieleBoley2021}

Early in this new era, it became apparent that a rapid rise in the number of objects orbiting the Earth, given certain of their physical characteristics, would in one way or another interfere with the cosmic light that some observers on the ground were accustomed to seeing.~\citep{McDowell2020} Professional astronomers saw their data impacted negatively by the bright trails of satellites reflecting sunlight to the Earth.~\citep{Zhang2020,Pularova2022} At least one instance has been reported in which the discovery of a faint, apparently transient object thought to represent an exotic astrophysical phenomenon~\citep{Jiang2020} was later suggested to have been caused instead by sunlight glinting from a satellite or space debris.~\citep{Nir2021,Michalowski2021} Astrophotographers experienced a surge in the number of their images affected by trails; in some cases, the images were not recoverable in software.~\citep{McFallJohnsen2020} People around the world for whom the night sky is an object of cultural and/or religious significance saw unwelcome reminders of changes imposed on them about which they were not consulted.~\citep{Noon2022} The future this implied was labeled `dystopian’ and `unsustainable’ by some observers.~\citep{Resnick2020,Witze2022}

Even considering the dawn of the Space Age some 65 years ago and the visibility of satellites in the night sky during the intervening decades, there is no historical precedent for the ongoing changes to how near-Earth space is used by humans. The existing international space policy framework dates to the mid-1960s and envisions a world that does not much resemble today with respect to how space is actually used.~\citep{Kazlouskaya2021} A few international bodies such as the International Telecommunication Union coordinate allocation of satellite orbits  as a kind of outer space equivalent of air-traffic control. Meanwhile, space becomes increasingly militarized by states, and some among them acting in bad faith carry out weapons tests that threaten all countries’ space assets.~\citep{Balmforth2021}

Against the backdrop of rapid technical achievement and advancement, a frontier/colonizing mentality among some actors, rising potential for military conflict in space, and an orbital realm becoming more crowded by the day, some have called for a new sense of `space environmentalism'.~\citep{Lawrence2022} This view aims to accommodate the right of all to access space while ensuring its development proceeds along ethical, sustainable and safety-conscious lines. At the opposite end of the spectrum is a sort of `space manifest destiny’ in which the first to arrive wins whatever resources they intend to exploit,~\citep{Billings1997} a framing that some governments have reinforced with their rhetoric.~\citep{Koren2020}

In 2020, an effort led predominantly by the professional astronomy community sought to open a dialogue with both the private space industry and its regulators on the topic of the responsible development of commercial space in a way that did not negatively affect ground-based astronomy. In a series of four international conferences in 2020-21, participants noted that there were many affected groups not represented in the process. While urging deliberate efforts to solicit their participation in future meetings, they also recognized the need to centralize and formalize the effort so as to carry forward the sense of collaboration initiated in the conferences. The establishment of the International Astronomical Union Centre for the Protection of the Dark and Quiet Sky from Satellite Constellation Interference (CPS) in 2022 is a concrete result of the conferences.~\citep{IAU2022} To fulfill the goal of increasing participation by affected stakeholder groups, CPS identified Community Engagement (CE) as one of four `hubs’ of activity. ``CE Hub'' is the official CPS outlet for engaging in broad stakeholder engagement.

The purpose of this report is to outline a framework guiding CE Hub as it begins its existence. After a more detailed review of the history leading up to the establishment of CPS, we reiterate and comment on the recommendations for improving stakeholder engagement. Finally, we present an initial strategy for implementing the recommendations that will guide the work of CE Hub in coming years.


\section{The Pre-CPS Community Engagement Landscape}

\subsection{Background and Context}

The notion of satellites and space debris (to which we refer collectively as ``space objects'') impacting the visibility of the night sky, and affecting astronomical observations, is not new. Concerns arose soon after the earliest satellite launches, and the effects on astronomy were reviewed multiple times in the intervening decades.~\citep{Cohen1989,Kovalesky1992,McNally1997,McNally1999,PortreeLoftus1999} Astronomers developed methods for mitigating adverse effects in their data after digital imaging was first deployed in the 1980s, although satellite effects masquerading as real astrophysical phenomena persisted.~\citep{Schaefer1987} There is little evidence in either the historical record or academic literature for high-level contacts between professional astronomers and organizations that launched and operated satellites to address the problem. 

Furthermore, there is a lengthy history of the developers of technology failing to consult with affected stakeholders prior to deploying those technologies.~\citep{Nye2006} Affected communities are often left on their own to deal with adverse impacts, while the benefits largely accrue to those who create that adversity.~\citep{Stilgoe2013} Scholars have gradually come to understand this history and to advocate for responsible technology development in the future. In terms of uses of near-Earth space, it has been argued that the existing international space policy regime may be inadequate to confront the rapidly changing situation involving large satellite constellations.~\citep{Barentine2022}

\subsection{Arrival of the `Large Constellation’ Era}

Small satellite constellations consisting of up to ~100 objects were launched beginning in the 1970s. By the 1990s, engineers realized that large flotillas of satellites in LEO could provide low-latency telecommunications with global coverage.~\citep{Maral1991,Crosbie1993} The barrier to this kind of use of near-Earth space was largely of an economic nature, given the historically high cost of launch. Technological progress involving hardware miniaturization and the advent of significant private investment in commercial spaceflight drastically reduced the cost of launch throughout the 2010s; by 2020, the launch cost per kilogram of payload was a factor of twenty lower than a decade earlier.~\citep{Jones2018}

In May 2019, Space Exploration Technologies Corp. (`SpaceX’) launched the first 60 satellites in its `Starlink’ constellation.~\citep{Chang2019} The project has the goal of providing uninterrupted, global Internet access from LEO with low latency, and the U.S. Federal Communications Commission (FCC) has licensed SpaceX to launch up to 42,000 satellites.~\citep{Henry2019} As of mid-2023, about 4,600 Starlink satellites have been launched;~\citep{McDowell2023} this number exceeds the total number of all functional satellites in orbit around the Earth before the first Starlink launch by about a factor of two.

There are many users of the night sky beyond professional astronomers.~\footnote{Some communities and individuals object to the concept of ``using'' of the night sky (as well as the space environment), in that such language could be interpreted to imply that these are simply ``objects'' that are at humans’ disposal to use however they wish. Instead, they see the night sky (and the Earth as well) as environments with which one is in relationship with. Many words have connotations that can imply assumptions that not all people share — something to be mindful of when engaging with diverse communities and perspectives.} These include avocational users, such as amateur astronomers, astrophotographers, and casual stargazers. Given the rapid development of `astrotourism’ as a form of sustainable recreation, and an industry supporting consumer demand for such experiences, there is an emerging professional class whose livelihood depends on preserving the natural qualities of the night sky. These activities may have significant rural economic development potential,~\citep{MitchellGallaway2019} which makes stakeholders of entire communities that benefit from the well-being of the astrotourism industry. They are also key to imparting the importance of appreciating and preserving the night sky.

Additionally, many people consider the night sky to be of cultural and/or religious significance, and access to the night sky may be seen as integral to their lives and well-being.~\citep{Gallaway2014,Blair2018} Their interest in the satellite constellation issue is one of definition, as connection with the night sky is an element of their concepts of self and community.~\citep{Blair2018} Many Indigenous communities, in particular, have extensive knowledge systems which involve ongoing relationship and interaction with both the Earth and sky.~\citep{Lee2016} The impacts of satellite constellations on the night sky for communities like these has not been considered in satellite constellation approval processes.  Members of Indigenous nations, as well as a variety of stakeholder groups have noted their historical exclusion from the process by which decisions that affect the quality of their night skies have been made.~\citep{SATCON2CEWG} Some Indigenous communities have benefitted from Starlink-provided internet because it enables connectivity where none had previously existed;~\citep{Duffy2020} however, some have argued that the constellations are a new and unwanted form of colonization.~\citep{Ferreira2021} Other authors challenge the assumption that satellite constellations are the only viable means of providing broadband to underserved populations.~\citep{Rawls2020}

\subsection{Response of the astronomical community}

Given their public reactions after the first Starlink launch, it is arguable that professional astronomers did not fully anticipate the threat that large satellite constellations might pose to their observations. Although satellites and space debris had already affected their work for decades, the potential for a sudden and significant rise in the severity of the interference was new. As they would later concede, the threat went largely unnoticed by the community as recently as 2020 when the National Academies’ decadal survey of U.S. astronomy and astrophysics was issued.~\citep{SATCON1}

Starlink satellites seen moving together in groups, some of which briefly became the brightest objects in the night sky, provoked quick responses from professional societies like the American Astronomical Society (AAS),~\citep{AAS2019} the Royal Astronomical Society (RAS),~\citep{RAS2019} and the International Astronomical Union (IAU).~\citep{IAU2019} Other organizations representing non-professional users of the night sky, such as the International Dark-Sky Association (IDA), issued their own statements.~\citep{IDA2019} Urgent consultation with SpaceX and other space operators was requested.

In conference with their members and affected stakeholders, some of these organizations arranged meetings of experts to assess the significance of the threat and evaluate possible solutions. The results were two series of two meetings each: the ``SATCON1'' and ``SATCON2'' conferences, convened by the National Science Foundation’s National Optical-Infrared Astronomy Research Laboratory with funding from AAS and the NSF, were U.S.-focused, while the Dark \& Quiet Skies for Science and Society conferences (``D\&QS-I'' and ``D\&QS-II''), convened by the IAU, the United Nations Office of Outer Space Affairs (UNOOSA) and the government of Spain (through the Instituto de Astrofísica de Canarias), had an international emphasis. The SATCONs were restricted in topical consideration to the satellite constellation issue only, while D\&QS also had a component about terrestrial skyglow.

\subsubsection{SATCON1 (29 June to 2 July 2020)}

SATCON1 was the first meeting of the four and dealt mostly with satellite impacts to professional astronomy, giving only limited consideration to ``Citizen Science, Amateur Astronomers, and Stargazers Worldwide''. It considered the scenarios for scope of impact from megaconstellations on these stakeholders, ranging from `minor’ to `severe’, noting that ``though we know of no plans to deploy a large unaided-eye-visible satellite constellation, there is no technical or legal barrier to building one.''\footnote{SATCON1 Report, ``Impacts on Scientific and Observational Programs'' \S9, p. 15.} 

Among its four Working Groups (WGs), the Metrics WG considered the ``Concerns of the non-professional astronomy community and adjacent night-sky stakeholders''.~\citep{SATCON1APX} It attempted to qualitatively assess potential impacts of planned large constellations on activities ranging from visual and telescopic observations of the night sky to various modes of astrophotography. Estimated impacts ran the gamut from `negligible’ to `fatal’. In reference to the other major technical recommendations of SATCON1, the Metrics WG wrote that ``ensuring that satellite visibility is unusual (relatively speaking) and that the time-averaged brightness of satellites is held below the threshold of naked-eye visibility is not a sufficient criterion to satisfy the needs of this constituency.'' This disconnect between the needs of professional astronomers and other users of the night sky set the stage for establishment of the Community Engagement Working Group in SATCON2.

\subsubsection{Dark \& Quiet Skies I (6-9 October 2020)}

D\&QS-I asserted that ``the cultural and religious practices of many people, especially those in indigenous societies, rely on access to the night sky''~\citep{DQSI} and further noted that the impacts of satellite constellations included potential effects on ``religious and cultural practices.''\footnote{Dark and Quiet Skies I report, \S2.5.1, p. 28.} It did not directly address community engagement, instead identifying ``observatories, industry, astronomy community, science funding agencies, national and international policymakers'' as key stakeholders.\footnote{\emph{Ibid}., \S2.5.2.4, p. 30.}

Recommendations made in the D\&QS-I report were forwarded to the United Nations Committee on the Peaceful Uses of Outer Space (UN-COPUOS) Scientific and Technical Subcommittee (STSC) at its 58th session in April 2021.\footnote{See COPUOS STSC Conference Room Paper A/AC.105/C.1/2021/CRP.17 (19 April 2021). \url{https://www.unoosa.org/oosa/oosadoc/data/documents/2021/aac.105c.12021crp/aac.105c.12021crp.17_0.html}} The proceedings included the following relevant statements:

\begin{quote}\small
[B]eyond science and technology, the pristine spectacle of the starry night sky has been inspirational to humankind since prehistoric times and this world cultural heritage should be zealously protected.\footnote{Dark and Quiet Skies I report, ``Introduction and Background'', \S3, pp. 1-2.}

A set of recommendations are proposed to ensure that the development of national space economies can continue with minimal interference to the science of astronomy. They are formulated within a framework of shared stewardship, in which all space stakeholders acknowledge and assume their responsibility to conserve Earth’s pristine natural night-time landscapes, grow national space industries and a global space economy, and support fundamental scientific research.\footnote{\emph{Ibid}., ``The Impact of Satellite Constellations on the Science of Astronomy'', \S20, p.~5.}

[I]t is proposed that IAU and UNOOSA engage with relevant stakeholders, intersessionally, on the matter of Dark and Quiet Skies, to form consensus on expert-recommendations and report back to COPUOS STSC 2022.\footnote{\emph{Ibid}., ``Conclusions'', \S139, p.~19.}
 \end{quote}

\subsubsection{SATCON 2 (12—16 July 2021)}

The main purpose of the second SATCON meeting was ``to develop specific, implementable paths to carrying out [the] recommendations'' of SATCON1.~\citep{SATCON2ES} However, it also aimed ``to engage a considerably wider group of stakeholders in the conversations than had been present at SATCON1.''\footnote{SATCON2 Executive Summary, p.~4.} Among the ``common themes'' of SATCON2 was ``the importance of continuing collaborative work between all stakeholders with highly diverse missions and motivations''.\footnote{\emph{Ibid}., p.~14.} The SATCON2 organizers specifically established a Community Engagement Working Group (CEWG), which brought many new voices and perspectives to the issue; its report contains many of the elements that we consider fundamental to the work of CE Hub.~\citep{Venkatesan2021}

\subsubsection{Dark \& Quiet Skies II (3-7 October 2021)}

The second Dark \& Quiet Skies meeting did not specifically address CE, but its report implies interest in the idea.~\citep{DQSIIWGR} The Satellite Constellation Working Group touched on the subject tangentially in \S8 (``Societal impacts''), and in particular, \S8.2 (``Dark skies and multiple stakeholders''): ``[It] is \ldots important to consider satellite constellation impacts in the context of unequal and underrepresented consequences for a diverse global range of societal stakeholders in the night sky.''\footnote{D\&QS-II report, \S8.2, p.~138.} In particular, D\&QS-II addressed the rights of Indigenous persons:
\begin{quote}\small
Indigenous Peoples, who possess deep and diverse cultural, religious, and historical relationships with the night sky, number among the global range of societal stakeholders invested in uncorrupted dark skies and should therefore be taken into consideration in accordance with these rights under international law.\footnote{\emph{Ibid}., ``Indigenous peoples' rights in international law'', \S8.2.2, p.~139.}
\end{quote}
The parties updated the STSC on the outcomes of D\&QS-II at its 59th session in February 2022.~\citep{UN2022}

\subsection{Establishment of CPS}

The D\&QS-I report suggested what eventually became CPS.\footnote{D\&QS-I report, recommendations R27-R30, pp.~154-55.} Its recommendations include:

\begin{quote}\small
\bf R27\rm: Organise professional development opportunities, collaborative research and exchanges of experience, including: developing proposals for conference sessions; submitting applications for collaboration networks, workshops and symposia; identifying possibilities for grant funding to develop competencies and knowledge.

\bf R28\rm: Develop techniques to include satellite constellation considerations in developing science cases for telescopes and instruments. Using a combination of predictive models of location and brightness, support the development of simulations to determine the broad impacts on science cases for existing and future observatories.

\bf R29\rm: Support an immediate coordinated effort for multiple spectral bands in optical and infrared observations of LEOsat constellation members, to characterize both slowly and rapidly varying reflectivity and the effectiveness of experimental mitigations.

\bf R30\rm: Support ``citizen science'' campaigns to involve the astrophotography community and amateur astronomy community in supporting data collection understanding of impacts.
\end{quote}

The Executive Committee of the IAU, in its 105th Meeting on 22 April 2021, approved the establishment of a ``Centre for the Protection of the Dark and Quiet Sky from Satellite Constellation Interference'' (IAU-CPS, or simply ``CPS''). A call for proposals was issued in June 2021, and the NOIRLab/SKAO proposal was selected in October 2021. CPS formally came into existence on 1 April 2022. In its announcement, IAU stated that
\begin{quote}\small
The CPS aims to become the leading voice for astronomical matters that relate to the protection of the dark and quiet sky from satellite constellations and to act as a hub of information and resources to which any stakeholder group, including the constellation satellite industry, will be able to contribute and from which they can draw in support of their own activities. The mission of the CPS is to coordinate efforts and unify voices across the global astronomical community with regard to the protection of the dark and quiet sky from satellite constellation interference.~\citep{IAU2022}
\end{quote}
To address this objective, CE Hub was created as part of CPS.


\section{Outcomes of SATCON2 and D\&QS-II}

In both the SATCON and Dark and Quiet Skies conference series, the second edition of each conference dealt in part with community engagement issues. We review the main findings and recommendations here. 

\subsection{Principal findings}

\noindent \bf All users of space and the night sky constitute the ``community''\rm. The D\&QS-II Satellite Constellation Working Group noted (echoing the language of the OST) that while ``general freedom exists to access space and that its exploration and use constitute provinces of all humankind, \ldots these freedoms are not absolute, and must find a limit in the freedoms exercised by other actors.''\footnote{D\&QS-II Satellite Constellation Working Group Report, ``Conclusions'', \S9, p.~139.} It further proceeded on the presumption that ``night skies for astronomy are a resource to all humanity, and the net of stakeholders in astronomical skies is far wider than only that of professional astronomy.''\footnote{\emph{Ibid}., ``Dark skies and access to science: heritage and cultural rights, and other stakeholders'', p.~137.} Among such additional stakeholders are amateur astronomers; those whose cultural or religious traditions are connected to celestial observations; and those who find that being able to view an unpolluted, natural night sky contributes positively to their overall sense of well-being. As noted in the CEWG report, ``a future transformation of the night sky [by large numbers of visible satellites] threatens to fundamentally rewrite the story of the relationship between humanity and the night sky.''\footnote{SATCON2 Community Engagement Working Group Report, p.~20.} \\

\noindent \bf Consideration of the diversity of stakeholders calls for direct and ongoing engagement of various individuals and groups\rm. SATCON2 identified ``the importance of continuing collaborative work between all stakeholders with highly diverse missions and motivations'' as among the `common themes’ of the meeting.\footnote{SATCON2 Executive Summary, p.~14.} Community engagement is key to bringing ``many new voices and perspectives to the issue''.\footnote{\emph{Ibid.}, p.~4.} The sense that these voices are ``new'' refers to the fact that many of these same stakeholders were effectively unseen in the past and left out of the dialogue among decision makers. We hold that they should be recognized and included in these discussions.\\

\noindent \bf There is a duty to consult with communities affected by satellite constellations, and it is especially important to engage with groups that have been historically left out of decisions having to do with uses of space and the night sky\rm.\footnote{SATCON2 Community Engagement Working Group Report, pp.~2-3.} The D\&QS-II Satellite Constellation Working Group recognized ``the applicability of indigenous rights protection in general international law, including relevant international declarations and treaties.''\footnote{D\&QS-II Satellite Constellation Working Group Report, \S8.2.2, p.~138.} The authors cited, \emph{inter alia}, the United Nations Declaration of the Rights of Indigenous Peoples;\footnote{UN General Assembly Resolution A/RES/61/295 (2007).} the United Nations International Labour Organization Convention No. 169;\footnote{Convention (No.~169) concerning indigenous and tribal people in independent countries, 27 June 1989, 1650 U.N.T.S. 383 (entered into force 5 September 1991).} and the Declaration adopted by the United Nations General Assembly of the World Conference on the Rights of Indigenous Peoples.\footnote{UN General Assembly Resolution A/RES/69/2 (22 September 2014).} As the authors noted, "The permanent loss of practicable scientific astronomy to global communities can therefore also affect heritage and cultural rights, which are protected under international law."\footnote{D\&QS-II Satellite Constellation Working Group Report \S8.2.2, p.~137.}\\

\noindent \bf It is important to be mindful of relevant treaties and laws pertaining to the rights of Indigenous Peoples\rm. As noted in the CEWG report, ``Indigenous peoples and nations must be consulted and their decisions should be respected.''\footnote{SATCON2 Community Engagement Working Group Report, p.~34.} Furthermore, the SATCON2 CEWG Report emphasized that ``Indigenous peoples in Canada and the United States are groups of sovereign nations with rights highlighted by treaties'' and that ``more work is needed for that discussion to be nation-to-nation and not colonizer-to-Indigenous peoples.''\footnote{\emph{Ibid.}, p.~33.}\\

\noindent \bf The megaconstellation phenomenon has caused people to think deeply, and perhaps for the first time, about what the rapid commercialization of near-Earth space means\rm. As noted in the D\&QS-II Satellite Constellation Working Group Report, communities are impacted that have little or even nothing to gain from the `New Space’ regime:

\begin{quote}\small
it is critical to refer to the plurality of interests among global and local stakeholders, in addition to States. This includes, \emph{inter alia}, a recognition of different sources of knowledge on and the significance of uninterrupted night skies, a shared global heritage of astronomy, navigation, and other practices including religious and cultural considerations. \ldots [S]atellite constellations will be seen by those in dark sky oases, by those in remote communities, and by those who may have little investment in space technology but maintain a long-standing cultural relationship with the stars.\footnote{D\&QS-II Satellite Constellation Working Group Report \S8.2 (``Dark skies and multiple stakeholders''), p.~137.}
\end{quote}

\noindent
Even within the professional astronomy community, the authors were concerned that \bf dealing with the burdens imposed by satellite constellations would fall disproportionately on those institutions and facilities with access to the least number of resources to deal with them\rm: ``The impacts of satellite constellations may also be felt inequitably across the global professional astronomical community. In practice, stakeholder conversations about available capacity to process corrupted data may be biased towards organizations with more resources, which is not necessarily reflective of the capabilities of the worldwide net of astronomical stakeholders.''\footnote{\emph{Ibid.}, \S8.2.1, p.~137.} Furthermore, the number of Indigenous professionals who may contribute to this work is remarkably low. Those few people are often tasked with representing the views of diverse communities, and their voices are easily drowned out. 

\subsection{Constituencies}

In understanding who the ``community'' is with whom we wish to engage, we considered the constituencies that it may comprise. First, we differentiated the ``community'' of CE from the ``community'' of users of near-Earth space as contemplated in Article I of the Outer Space Treaty (OST).\footnote{United Nations. (1967, January 27). Treaty on Principles Governing the Activities of States in the Exploration and Use of Outer Space, including the Moon and Other Celestial Bodies, 18 U.S.T. 2410, 610 U.N.T.S. 205, 61 I.L.M. 386. (hereafter the ``Outer Space Treaty'' or ``OST''}$^{,}$\footnote{``Outer space, including the moon and other celestial bodies, shall be free for exploration and use by all States without discrimination of any kind, on a basis of equality and in accordance with international law.'' (OST, Article I, Clause 2)} Second, we broadened the scope to include anyone with an interest in the terrestrial night sky as it is affected by uses (and users) of outer space. That scope considers groups not directly mentioned in the OST, including people who have no ambition to establish a presence in outer space. And third, we placed special emphasis on groups that have historically been left out of decision making about the uses of space and the night sky, in order to bring about more diverse participation and facilitate more equitable decision making processes.

We then considered the five specific constituencies ``that had not previously been explicitly included in SATCON1 or other policy discussions about satellite constellations, including some groups traditionally excluded from political and economic power''\footnote{SATCON2 Community Engagement Working Group Report, p.~4.} identified by the SATCON2 CEWG:

\begin{enumerate}
\item Astrophotography and Astro-Tourism
\item Amateur Astronomy
\item Indigenous Communities and Perspectives
\item Planetariums
\item Environmental and Ecological Concerns
\end{enumerate}

The D\&QS-II Satellite Constellation Working Group affirmed the identification of some of these constituencies, specifically amateur astronomers and astrophotographers\footnote{D\&QS-II Satellite Constellation Working Group Report \S8.2.1, p.~137.} and Indigenous persons.\footnote{\emph{Ibid.}, \S8.2.2, p.~138} The latter are ``among the global range of societal stakeholders invested in uncorrupted dark skies and should therefore be taken into consideration in accordance with these rights under international law.''\footnote{\emph{Ibid.}, p.~139.} It is also important to note an issue discussed in the SATCON2 CEWG report and noted by multiple Indigenous contributors: that as many Indigenous communities are sovereign nations, it is thus ``inappropriate'' for them to be discussed as a ``special interest group.''\footnote{SATCON2 Community Engagement Working Group Report, p.~33.}

The SATCON2 constituencies segmented along three major lines: (1) non-professional astronomers and users of the night sky for recreational and educational purposes; (2) marginalized social and cultural groups that are not often consulted in matters of the uses of technology as it impacts collective goods and commons; and (3) the broader environmental community beyond the dark-skies movement. The CEWG at the same time acknowledged ``that there remain many constituencies and perspectives not included in the Community Engagement Working Group that may prove important players in future negotiation and policy-making, such as telecommunication companies, space contractors, economic development groups, ground-based internet equipment suppliers, and Internet service providers.''\footnote{\emph{Ibid.}, p.~5.} This calls into question what defines ``community'' in the CPS context. For example, ``Industry'' has its own CPS hub. Here, ``industry'' seems to be understood to be the private commercial space industry, i.e., the companies that build, launch and operate satellites and constellations of satellites. To date it does not appear to include other groups. 

It is notable that the main business competitors of satellite broadband providers — namely, terrestrial broadband companies — were not consulted during the SATCON meetings at all. It may well be possible to provide broadband Internet access to billions of people through wireless telecommunications networks without the need for tens of thousands of satellites. This is particularly true in many developing economies, which have tended to ``leapfrog'' over intermediate technologies and proceed directly to high-speed wireless communications.~\citep{Singh1999}

Furthermore, it may not be possible to include all groups because the notion of ``community'' is still somewhat ill-defined. The SATCON2 CEWG found that the largest group of people not explicitly included in their Working Group was ``the population of humans world-wide who admire, cherish, view, connect with, seek solace from, practice traditional religion and culture with, navigate by, are inspired by, and need the stars, the Milky Way, and unpolluted night skies.''\footnote{SATCON2 Community Engagement Working Group Report, p.~5.} It is impossible to collect all such broad and diverse views, in the same sense that the groups with which we have had contact do not represent all of their members. The CEWG acknowledged this: ``Our compiled report does not speak for all members of any constituency, or all examples of a group, e.g., all Native American tribal communities or all environmental groups.'' \footnote{\emph{Ibid}.}

Indigenous communities and marginalized groups often experience a myriad of challenges. In 2020-21 they were ``overloaded by disproportionate fallout from climate change and the pandemic'',\footnote{\emph{Ibid}.} and thus a priori were not necessarily deeply engaged in the issues of SATCON1 and 2. That may improve in the future as the pandemic recedes, but it is important to be mindful that Indigenous communities may be facing other pressing social and environmental issues which may prevent them from being as engaged in discussing satellite constellation issues as they otherwise could be. 

Lastly, we recognize the reality of `crisis fatigue'. Setting the COVID-19 pandemic aside, humanity faces serious challenges now and in coming decades. It may be that some potential constituencies simply are not engaged in the topic of CPS’ work because time and effort are limited and they have to prioritize other concerns over this. While they may care about the large satellite constellation issue, on balance they may find that CPS is not where they want to invest their energy. We must keep in mind that this does not imply indifference any more than it means they do not believe that a problem exists.

\subsection{Common Themes and Principles}

The SATCON2 CWEG identified ``common themes that recurred and resonated''\footnote{\emph{Ibid}., p.~6.} with its constituencies:

\begin{quote}
\begin{enumerate}
\item The skies and space belong to everyone.\footnote{It is important to note that some readers may bristle at the notion that space and the night sky are here referred to in a possessive context. Some societies reject notions of human lordship or dominion over nature and the environment, and so may disagree with the implication that the sky and space are subject to human appropriation.} Space is a global commons.
\item All people are impacted by changes in the sky. Nearly all [who were] consulted for SATCON2 had already noticed a dramatic rise in satellite constellation sightings in the past two years, and were worried.
\item Many communities see the unchecked actions of space actors as colonization expanded to a cosmic scale during a time of global crisis.
\item The sky must be considered part of the environment and the current [U.S.] National Environmental Policy Act (NEPA) exemption for the satellite constellation industry must end.
\item Ecosystems depend on the night sky and on each other.
\end{enumerate}
\end{quote}

\noindent
These are not necessarily universal beliefs, nor are they in any way sacrosanct. SATCON2 was a data point, but it’s far from the only one and it’s not by any means the last word on the subject. It is also the only sample we have to date, so we expect to hear these and other perspectives from other groups with which we establish contact in coming years. 

Things have changed even in the two years that have passed since SATCON2. There are now many more satellites, and the pace of launches continues to increase. However, unlike in May 2019, the appearance of large groups of recently launched satellites in the night sky is no longer a novel phenomenon for many. In large part, the shock of that early period has worn off, and the media are covering the story with much less frequency. We recognize the possibility that this issue may not currently be one of ongoing public interest, and also that it may be difficult to raise and sustain that interest in the future.

\subsection{Recommendations}

Overall, we agree with the assertion in the Dark \& Quiet Skies report that ``there is a compelling need to formalize engagement on the impact of satellite constellations.''\footnote{D\&QS-II Working Group Reports, \S4.7 (``Involving Citizens''), p.~234.} In addition, ``it is critical to refer to the plurality of interests among global and local stakeholders located in diverse places around the world, in addition to States.''\footnote{\emph{Ibid}., \S8.2 (``Dark skies and multiple stakeholders''), p.~137.} This ethos entails establishing contact with a true diversity of constituencies, among them individuals, organizations and governments. 

The SATCON2 CEWG summarized the findings of its subgroups with a series of broad ``recommendations to decisionmakers, regulators, the satellite industry, researchers, and all communities affected by satellite constellations.''\footnote{SATCON2 Community Engagement Working Group Report, p.~6.} We restate here the recommendations we find relevant to the mission of CE Hub below along with our interpretations of their meanings for our work.\footnote{We determined that two additional recommendations, ``Better international regulation and globally coordinated oversight/enforcement'' and ``Slow or stop satellite constellation launches until problems are resolve'', are beyond the remit of the CE Hub and we do not comment on them here.}

\begin{enumerate}
\item \bf Duty to consult\rm. This proceeds from a position that ``Space belongs to us all and we need to listen to all constituencies impacted by satellite constellations.''\footnote{SATCON2 Community Engagement Working Group Report, \S1.5.1 (``Duty to consult''), p.~75. We preserve the language in this direct quotation while noting the previously expressed concern about framing the night sky and space in terms of words like ``belong'', which implies human ownership of, or the right to exploit, nature.}  Responsible use of any commons comes with an obligation to seek and understand the views of parties expected to be impacted by activities affecting it.
\item \bf Need for more information and communication\rm. Proceeding from a belief that knowledge is power, stakeholders need access to timely and accurate information about satellite constellations. While it is not our role to interpret that information or suggest its uses to our constituencies, we feel a responsibility to serve to the best of our abilities as a conduit for information and to create conditions where meaningful exchanges of ideas can take place. 
\item \bf Engage with industry\rm. Noting that the private commercial space industry is a legitimate user of space under the OST, we agree that any effort to engage the community of stakeholders described here necessarily entails respectful dialogue with space companies and other private space actors. 
\item \bf Recognize and rebalance power structures\rm. We agree with the D\&QS-II observation that while ``general freedom exists to access space and that its exploration and use constitute provinces of all humankind, \ldots these freedoms are not absolute, and must find a limit in the freedoms exercised by other actors.''\footnote{``-II Satellite Constellation Working Group Report, \S3 (``Recommendations for the international governance of outer space''), p.~83.} Those `other actors’ sometimes behave in ways that yield negative externalities that particularly affect societies that have generally been shut out of receiving many of the benefits of technology development throughout history. Those same societies deserve consideration of their interests in using the shared resource of outer space. We shall therefore, quoting the language of the SATCON2 CWEG, strive to create a model of ``broad inclusion of all affected communities in meaningful dialogue from the start.''\footnote{SATCON2 Community Engagement Working Group Report, \S1.5.4 (``Recognize and rebalance power structures''), p.~76.} Direct engagement and consultation is the means by which we intend to build respectful relationships to serve the interest of meaningful interactions.
\item \bf Learn from the past\rm. History is replete with examples of disruptive technologies that developed largely or entirely in the absence of regulation, some of which led to significant global challenges. In some cases, these developments have led to the loss of traditional knowledge and ways of life, leaving legacies of colonization and social inequities. We aim to learn from history in hopes of avoiding its repetition.
\item \bf ``Science vs. Internet'' is a false choice\rm. We recognize the tremendous potential of broadband Internet access to improve standards of living. At the same time, we reject the notion that realizing that potential creates an unsolvable problem pitting the consequences of infrastructure development against access to a resource that could change the lives of billions of people. There are no ``sides'' to choose in this debate, because the protection of dark and quiet skies and the deployment of large-scale broadband Internet access are not mutually exclusive objectives.
\item \bf Continued active engagement and conversation\rm. We understand that pursuit of the CE Hub’s mission is perpetual, and that given the pace of technological development, it is unlikely to be achieved in the near-term. We therefore commit to the CE Hub of CPS as an activity that will require resources into the foreseeable future.
\end{enumerate}


\section{Outlook and Future Plans}

In the run-up to the formal establishment of CPS, we represented that the CE Hub activity would proceed initially as follows:\footnote{Quoting the CE Hub page on the CPS website (\url{https://cps.iau.org/community-engagement-hub/}).}

The CE activity will begin by describing the study case on the basis of the findings contained in the above-mentioned Reports in a language that can be understood by the widest possible communities. In this activity, the CE could benefit from the collaboration with the IAU Division C (Education, Outreach and Heritage), Commission C4 (World Heritage and Astronomy), Commission B7 (Protection of Existing and Potential Observatory Sites) and Office of Astronomy Outreach (OAO) and its Network of national interfaces.

In a second phase, the CE will engage directly a variety of representative cultural communities in collecting their position on the case study and their suggestions on addressing the more general theme of the balance between technological advancement and protection of the environment and of human cultural heritage. It will solicit input through interviews, discussion fora, town halls and similar events and activities intended to provide spaces for civil discussions on all topics within the purview of CPS and its mission. The outputs from this effort may include recordings of webinars and other online events; white papers and reports; and resources stakeholders can use to advance the conversation within their own communities.

CE will further seek the expertise of those in the environmental history and Science and Technology Studies (STS) communities in order to better understand the past history of such interactions between technology and society. In collaboration with stakeholders, we will seek to articulate their views through the lens of history, including colonialism, the exploitation of marginalized societies and groups, and ongoing efforts to achieve social and environmental justice.

In our Hub activities, we will strive for independence and objectivity. Our task is not ``advocacy'', in which a particular point of view or opinion is promoted. To the extent we are advocates, our advocacy is on behalf of our stakeholder community in soliciting, recording and sharing of its diverse views on the topics within the purview of CPS. We envision the hub as a conduit or channel for dialog and our role as moderators whose task is not to select ``winners and losers'', but rather to ensure a fair and respectful exchange of ideas. 

This report represents the first element of the plan. In parallel, we will compile a list of preliminary stakeholder groups and establish contact with them. We also intend to reach out to community engagement experts and practitioners to better understand best practices. 

\subsection{External communications}

This activity aims to broaden the participation in discussions about satellite constellations by actively reaching out to stakeholders and communities that have not participated to date. It involves cultivating relationships with members of these groups, and others already involved, with the goal of continuing to solicit input as the satellite constellation issue evolves in coming years. We will provide timely updates on satellite constellation issues to stakeholders and the public and help raise awareness of the CPS and IAU’s efforts around the issue of dark and quiet skies.

\subsection{Cooperation within and beyond IAU}

IAU has for many years developed expertise in the areas of outreach, education and development, manifested in its Offices of Astronomy Outreach (OAO), Astronomy for Education (OAE) and Astronomy for Development (OAD). The experiences of these offices and their staff are invaluable guides to our work. We anticipate direct interface with each of these offices to learn from their experiences and identify opportunities for collaboration. This will better inform our activities and avoid duplication of effort. We will further seek cooperation with organizations external to IAU sharing its interests and goals, and which may also benefit CE Hub’s activities and objectives.

\subsection{Soliciting perspectives and fostering dialogue}

We plan to gather a broad array of stakeholder views on satellite constellation issues, and  share the findings with CPS and the general public via reports and white papers. In addition, we will convene online and/or in-person events such as small group discussions and webinars that create a space for both one-way communication of stakeholder views as well as dialogues among them. For some events, we may solicit guest topical speakers/presenters and have time for discussion afterwards. 

\subsection{Building relationships}

In our work in CE, we will strive to help facilitate the building of meaningful relationships between various parties participating in and affected by issues pertaining to satellite constellations.  We seek to provide opportunities for people and communities coming from diverse places and points of view to share their perspectives in a respectful environment.  We realize that it is possible that such interactions may sometimes result in individuals and communities simply agreeing to disagree, but it is our hope that through the work of our Hub, we can help bring about increased communication, awareness and collaborative endeavors among those who wish to work together to find meaningful solutions to emerging issues related to satellite constellations and the night sky.


\section{Acknowledgments}

The authors wish to acknowledge the SATCON2 Community Engagement Working Group for their efforts in devising the framework used to create this report. They also extend special thanks to Juan Carlos Chavez (Founder, Cosmic Deer Dancer, LLC) for a thoughtful review of the manuscript.\\

This work is licensed under the Creative Commons Attribution-ShareAlike 4.0 International License. To view a copy of this license, visit \url{http://creativecommons.org/licenses/by-sa/4.0/}.


\bibliography{foundational-report}

\begin{thebibliography}{53}
\providecommand{\natexlab}[1]{#1}
\providecommand{\url}[1]{\texttt{#1}}
\providecommand{\urlprefix}{URL }
\expandafter\ifx\csname urlstyle\endcsname\relax
  \providecommand{\doi}[1]{DOI:\discretionary{}{}{}#1}\else
  \providecommand{\doi}{DOI:\discretionary{}{}{}\begingroup
  \urlstyle{rm}\Url}\fi

\bibitem[{{AAS}(2019)}]{AAS2019}
{AAS} (2019) {AAS} issues position statement on satellite constellations (10
  {June} 2019).
\newblock
  \urlprefix\url{https://aas.org/press/aas-issues-position-statement-satellite-constellations}.

\bibitem[{Balmforth(2021)}]{Balmforth2021}
Balmforth T (2021) 'razor-sharp precision': Russia hails anti-satellite weapons
  test (17 {November} 2021).
\newblock \emph{Reuters}
  \urlprefix\url{https://www.reuters.com/business/aerospacedefense/russia-dismisses-us-criticism-anti-satellite-weapons-test-2021-11-16/}.

\bibitem[{Barentine et~al.(2022)Barentine, Heim, Venkatesan, Lowenthal and
  Vidaurri}]{Barentine2022}
Barentine J, Heim J, Venkatesan A, Lowenthal J and Vidaurri M (2022)
  Reimagining near-earth space policy in a post-covid world.
\newblock \emph{Virginia Policy Review} 15(1): 58--86.
\newblock \doi{10.5281/ZENODO.6903582}.
\newblock \urlprefix\url{https://zenodo.org/record/6903582}.

\bibitem[{Billings(1997)}]{Billings1997}
Billings L (1997) Frontier days in space: are they over?
\newblock \emph{Space Policy} 13(3): 187--190.
\newblock \doi{10.1016/s0265-9646(97)00020-9}.
\newblock \urlprefix\url{https://doi.org/10.1016/s0265-9646(97)00020-9}.

\bibitem[{Blair(2018)}]{Blair2018}
Blair A (2018) An exploration of the role that the night sky plays in the lives
  of the dark sky island community of sark.
\newblock \emph{Journal of Skyscape Archaeology} 3(2): 236--252.
\newblock \doi{10.1558/jsa.34689}.
\newblock \urlprefix\url{https://doi.org/10.1558/jsa.34689}.

\bibitem[{Boley and Byers(2021)}]{BoleyByers2021}
Boley AC and Byers M (2021) Satellite mega-constellations create risks in low
  earth orbit, the atmosphere and on earth.
\newblock \emph{Scientific Reports} 11(1).
\newblock \doi{10.1038/s41598-021-89909-7}.
\newblock \urlprefix\url{https://doi.org/10.1038/s41598-021-89909-7}.

\bibitem[{Chang(2019)}]{Chang2019}
Chang K (2019) {SpaceX} launches 60 starlink internet satellites into orbit (23
  {May} 2019).
\newblock \emph{New York Times}
  \urlprefix\url{https://www.nytimes.com/2019/05/23/science/spacex-launch.html}.

\bibitem[{Cohen(1989)}]{Cohen1989}
Cohen R (1989) The threat to radio astronomy from radio pollution.
\newblock \emph{Space Policy} 5(2): 91--93.
\newblock \doi{10.1016/0265-9646(89)90064-7}.
\newblock \urlprefix\url{https://doi.org/10.1016/0265-9646(89)90064-7}.

\bibitem[{Crosbie(1993)}]{Crosbie1993}
Crosbie DB (1993) The new space race satellite mobile communications.
\newblock \emph{{IEE} Review} 39(3): 111.
\newblock \doi{10.1049/ir:19930054}.
\newblock \urlprefix\url{https://doi.org/10.1049/ir:19930054}.

\bibitem[{Duffy(2020)}]{Duffy2020}
Duffy K (2020) How {SpaceX} teamed up with a small canadian it company to bring
  its starlink internet service to an indigenous community (19 {December}
  2020).
\newblock \emph{Business Insider}
  \urlprefix\url{https://www.businessinsider.com/spacex-starlink-internet-first-canada-customer-indigenous-community-pikangikum-musk-2020-12}.

\bibitem[{Ferreira(2021)}]{Ferreira2021}
Ferreira R (2021) {SpaceX}'s satellite megaconstellations are astrocolonialism,
  indigenous advocates say (5 {October} 2021).
\newblock \emph{Vice Motherboard}
  \urlprefix\url{https://www.vice.com/en/article/k78mnz/spacexs-satellite-megaconstellations-are-astrocolonialism-indigenous-advocates-say}.

\bibitem[{Gallaway(2014)}]{Gallaway2014}
Gallaway T (2014) The value of the night sky.
\newblock In: Meier J, Hasen{\"o}hrl U, Krause K and Pottharst M (eds.)
  \emph{Urban Lighting, Light Pollution and Society}, chapter~14. New York:
  Routledge, pp. 267--283.

\bibitem[{Hall et~al.(2021)Hall, Walker, Rawls, McDowell, Seaman, Venkatesan,
  Lowenthal, Green, Krafton and Parriott}]{SATCON2ES}
Hall J, Walker C, Rawls M, McDowell J, Seaman R, Venkatesan A, Lowenthal J,
  Green R, Krafton K and Parriott J (2021) Executive {Summary} of the {SATCON2}
  {Workshop}.
\newblock Technical Report 10.3847/25c2cfeb.4554c01f, NSF's NOIRLab.
\newblock
  \urlprefix\url{https://noirlab.edu/public/media/archives/techdocs/pdf/techdoc031.pdf}.

\bibitem[{Henry(2019)}]{Henry2019}
Henry C (2019) {SpaceX} submits paperwork for 30,000 more starlink satellites
  (15 {October} 2019).
\newblock \emph{SpaceNews}
  \urlprefix\url{https://spacenews.com/spacex-submits-paperwork-for-30000-more-starlink-satellites/}.

\bibitem[{{IAU}(2019)}]{IAU2019}
{IAU} (2019) {IAU} statement on satellite constellations (3 {June} 2019).
\newblock
  \urlprefix\url{https://www.iau.org/news/announcements/detail/ann19035/}.

\bibitem[{{IAU}(2021)}]{DQSI}
{IAU} (2021) Dark and {Quiet} {Skies} for {Science} and {Society} {On}-line
  {Workshop}: {Report} and recommendations.
\newblock Technical Report 10.5281/zenodo.5898785, International Astronomical
  Union.
\newblock
  \urlprefix\url{https://www.iau.org/static/publications/dqskies-book-29-12-20.pdf}.

\bibitem[{{IAU}(2022)}]{IAU2022}
{IAU} (2022) Launch of {New} {IAU} {Centre} {Safeguarding} {Astronomy} from
  {Satellite} {Constellation} {Interference} (10 {June} 2022).
\newblock
  \urlprefix\url{https://www.iau.org/news/announcements/detail/ann22024/}.

\bibitem[{{IDA}(2019)}]{IDA2019}
{IDA} (2019) Response to {SpaceX} {Starlink} {Low} {Earth} {Orbit} {Satellite}
  {Constellation} (29 {May} 2019).
\newblock \urlprefix\url{https://www.darksky.org/starlink-response/}.

\bibitem[{Jiang et~al.(2020)Jiang, Wang, Zhang, Kashikawa, Ho, Cai, Egami,
  Walth, Yang, Zhang and Zhao}]{Jiang2020}
Jiang L, Wang S, Zhang B, Kashikawa N, Ho LC, Cai Z, Egami E, Walth G, Yang YS,
  Zhang BB and Zhao HB (2020) A possible bright ultraviolet flash from a galaxy
  at redshift z{\hspace{0.167em}}$\approx${\hspace{0.167em}}11.
\newblock \emph{Nature Astronomy} 5(3): 262--267.
\newblock \doi{10.1038/s41550-020-01266-z}.
\newblock \urlprefix\url{https://doi.org/10.1038/s41550-020-01266-z}.

\bibitem[{Jones(2018)}]{Jones2018}
Jones HW (2018) The recent large reduction in space launch cost.
\newblock In: \emph{48th International Conference on Environmental Systems,
  Albuquerque, New Mexico, USA, 8-12 July 2018}, ICES-2018-81.
\newblock \urlprefix\url{https://ttu-ir.tdl.org/handle/2346/74082}.

\bibitem[{Kazlouskaya(2021)}]{Kazlouskaya2021}
Kazlouskaya M (2021) Large satellite constellations: Legal challenges in
  addressing space sustainability and astronomical observations.
\newblock \emph{German Journal of Air and Space Law} 70(571).

\bibitem[{Koren(2020)}]{Koren2020}
Koren M (2020) No one should `colonize' space (17 {September} 2020).
\newblock \emph{The Atlantic}
  \urlprefix\url{https://www.theatlantic.com/science/archive/2020/09/manifest-destiny-trump-space-exploration/612439/}.

\bibitem[{Kovalesky(1992)}]{Kovalesky1992}
Kovalesky J (1992) Satellites, space debris, aircraft and astronomy.
\newblock In: \emph{NATO Committee on the Challenges of Modern Society}.
  Gif-sur-Yvette, France: Atlantica S{\'e}guier Fronti{\`e}res, pp. 143--158.

\bibitem[{Lawrence et~al.(2022)Lawrence, Rawls, Jah, Boley, {Di Vruno},
  Garrington, Kramer, Lawler, Lowenthal, McDowell and
  McCaughrean}]{Lawrence2022}
Lawrence A, Rawls ML, Jah M, Boley A, {Di Vruno} F, Garrington S, Kramer M,
  Lawler S, Lowenthal J, McDowell J and McCaughrean M (2022) The case for space
  environmentalism.
\newblock \emph{Nature Astronomy} 6(4): 428--435.
\newblock \doi{10.1038/s41550-022-01655-6}.
\newblock \urlprefix\url{https://doi.org/10.1038/s41550-022-01655-6}.

\bibitem[{Lee(2016)}]{Lee2016}
Lee AS (2016) Ojibwe {Giizhiig} {Anung} {Masinaaigan} and {D(L)akota}
  {Mako{\.c}e} {Wi{\.c}aŋ{\.h}pi} {Wowapi}: {Revitalization} of native
  american star knowledge, a community effort.
\newblock \emph{Journal of Astronomy in Culture} 1(1): 41--56.
\newblock \urlprefix\url{https://escholarship.org/uc/item/58m4f9pq}.

\bibitem[{Maral et~al.(1991)Maral, de~Ridder, Evans and Richharia}]{Maral1991}
Maral G, de~Ridder JJ, Evans BG and Richharia M (1991) Low earth orbit
  satellite systems for communications.
\newblock \emph{International Journal of Satellite Communications} 9(4):
  209--225.
\newblock \doi{10.1002/sat.4600090403}.
\newblock \urlprefix\url{https://doi.org/10.1002/sat.4600090403}.

\bibitem[{McDowell(2020)}]{McDowell2020}
McDowell JC (2020) The low earth orbit satellite population and impacts of the
  {SpaceX} starlink constellation.
\newblock \emph{The Astrophysical Journal} 892(2): L36.
\newblock \doi{10.3847/2041-8213/ab8016}.
\newblock \urlprefix\url{https://doi.org/10.3847/2041-8213/ab8016}.

\bibitem[{McDowell(2023)}]{McDowell2023}
McDowell JC (2023) Jonathan's space pages - starlink statistics.
\newblock \urlprefix\url{https://planet4589.org/space/con/star/stats.html}.

\bibitem[{McFall-Johnsen(2020)}]{McFallJohnsen2020}
McFall-Johnsen M (2020) Elon musk's starlink satellites photobombed comet
  neowise in a photographer's striking image (29 {July} 2020).
\newblock \emph{Business Insider}
  \urlprefix\url{https://www.businessinsider.com/elon-musk-starlink-satellites-photobomb-comet-neowise-2020-7}.

\bibitem[{McNally(1997)}]{McNally1997}
McNally D (1997) Adverse effects of space debris on astronomy.
\newblock \emph{Advances in Space Research} 19(2): 399--402.
\newblock \doi{10.1016/s0273-1177(97)00020-3}.
\newblock \urlprefix\url{https://doi.org/10.1016/s0273-1177(97)00020-3}.

\bibitem[{McNally and Rast(1999)}]{McNally1999}
McNally D and Rast R (1999) The effect of spacecraft and space debris on
  astronomical observation.
\newblock \emph{Advances in Space Research} 23(1): 255--258.
\newblock \doi{10.1016/s0273-1177(99)00011-3}.
\newblock \urlprefix\url{https://doi.org/10.1016/s0273-1177(99)00011-3}.

\bibitem[{Micha{\l}owski et~al.(2021)Micha{\l}owski, Kami{\'{n}}ski,
  Kami{\'{n}}ska and Wnuk}]{Michalowski2021}
Micha{\l}owski MJ, Kami{\'{n}}ski K, Kami{\'{n}}ska MK and Wnuk E (2021)
  {GN}-z11-flash from a man-made satellite not a gamma-ray burst at redshift
  11.
\newblock \emph{Nature Astronomy} 5(10): 995--997.
\newblock \doi{10.1038/s41550-021-01472-3}.
\newblock \urlprefix\url{https://doi.org/10.1038/s41550-021-01472-3}.

\bibitem[{Mitchell and Gallaway(2019)}]{MitchellGallaway2019}
Mitchell D and Gallaway T (2019) Dark sky tourism: economic impacts on the
  colorado plateau economy, {USA}.
\newblock \emph{Tourism Review} 74(4): 930--942.
\newblock \doi{10.1108/tr-10-2018-0146}.
\newblock \urlprefix\url{https://doi.org/10.1108/tr-10-2018-0146}.

\bibitem[{Nir et~al.(2021)Nir, Ofek and Gal-Yam}]{Nir2021}
Nir G, Ofek EO and Gal-Yam A (2021) The {GN}-z11-flash event can be a satellite
  glint.
\newblock \emph{Research Notes of the {AAS}} 5(2): 27.
\newblock \doi{10.3847/2515-5172/abe540}.
\newblock \urlprefix\url{https://doi.org/10.3847/2515-5172/abe540}.

\bibitem[{Noon(2022)}]{Noon2022}
Noon K (2022) Thousands of satellites are polluting australian skies, and
  threatening ancient indigenous astronomy practices (19 {April} 2022).
\newblock \emph{The Conversation}
  \urlprefix\url{https://theconversation.com/thousands-of-satellites-are-polluting-australian-skies-and-threatening-ancient-indigenous-astronomy-practices-173840}.

\bibitem[{Nye(2006)}]{Nye2006}
Nye D (2006) \emph{Technology matters: Questions to live with}.
\newblock Cambridge, Massachusetts: MIT Press.

\bibitem[{Portree and Loftus(1999)}]{PortreeLoftus1999}
Portree D and Loftus J (1999) Orbital debris: A chronology.
\newblock NASA Technical Paper 208856, National Aeronautics and Space
  Administration.

\bibitem[{Pultarova(2022)}]{Pularova2022}
Pultarova T (2022) {SpaceX}'s starlink satellites leave streaks in
  asteroid-hunting telescope's images (20 {January} 2022).
\newblock \emph{Scientific American}
  \urlprefix\url{https://www.scientificamerican.com/article/spacexs-starlink-satellites-leave-streaks-in-asteroid-hunting-telescopes-images/}.

\bibitem[{{RAS}(2019)}]{RAS2019}
{RAS} (2019) {RAS} statement on starlink satellite constellation (7 {June}
  2019).
\newblock
  \urlprefix\url{https://ras.ac.uk/news-and-press/news/ras-statement-starlink-satellite-constellation}.

\bibitem[{Rawls et~al.(2020)Rawls, Thiemann, Chemin, Walkowicz, Peel and
  Grange}]{Rawls2020}
Rawls ML, Thiemann HB, Chemin V, Walkowicz L, Peel MW and Grange YG (2020)
  Satellite constellation internet affordability and need.
\newblock \emph{Research Notes of the {AAS}} 4(10): 189.
\newblock \doi{10.3847/2515-5172/abc48e}.
\newblock \urlprefix\url{https://doi.org/10.3847/2515-5172/abc48e}.

\bibitem[{Resnick(2020)}]{Resnick2020}
Resnick B (2020) The night sky is increasingly dystopian (29 {January} 2020).
\newblock \emph{Vox}
  \urlprefix\url{https://www.vox.com/science-and-health/2020/1/7/21003272/space-x-starlink-astronomy-light-pollution}.

\bibitem[{Schaefer et~al.(1987)Schaefer, Barber, Brooks, Deforrest, Maley,
  Norman~W., McNiel, Noymer, Presnell, Schwartz and Whitney}]{Schaefer1987}
Schaefer BE, Barber M, Brooks JJ, Deforrest A, Maley PD, Norman~W IM, McNiel R,
  Noymer AJ, Presnell AK, Schwartz R and Whitney S (1987) The perseus flasher
  and satellite glints.
\newblock \emph{The Astrophysical Journal} 320: 398.
\newblock \doi{10.1086/165552}.
\newblock \urlprefix\url{https://doi.org/10.1086/165552}.

\bibitem[{Singh(1999)}]{Singh1999}
Singh JP (1999) \emph{Leapfrogging Development? The Political Economy of
  Telecommunications Restructuring}.
\newblock New York: State University of New York Press.

\bibitem[{Stilgoe et~al.(2013)Stilgoe, Owen and Macnaghten}]{Stilgoe2013}
Stilgoe J, Owen R and Macnaghten P (2013) Developing a framework for
  responsible innovation.
\newblock \emph{Research Policy} 42(9): 1568--1580.
\newblock \doi{10.1016/j.respol.2013.05.008}.
\newblock \urlprefix\url{https://doi.org/10.1016/j.respol.2013.05.008}.

\bibitem[{Thiele and Boley(2021)}]{ThieleBoley2021}
Thiele S and Boley AC (2021) Investigating the risks of debris-generating asat
  tests in the presence of megaconstellations.
\newblock \doi{10.48550/ARXIV.2111.12196}.
\newblock \urlprefix\url{https://arxiv.org/abs/2111.12196}.

\bibitem[{{UN}(2022)}]{UN2022}
{UN} (2022) Protection of {Dark} and {Quiet} {Skies}.
\newblock Working Paper A/AC.105/C.1/L.396, Committee on the Peaceful Uses of
  Outer Space Science and Technology Subcommittee, Vienna.
\newblock
  \urlprefix\url{https://www.unoosa.org/res/oosadoc/data/documents/2022/aac_105c_1l/aac_105c_1l_396_0_html/AC105_C1_L396E.pdf}.

\bibitem[{Venkatesan et~al.(2021{\natexlab{a}})Venkatesan, Lowenthal, Arion,
  Avila~Castro, Bannister, Barentine, Begay, Chavez, Carttar, Gering, Hartley,
  Hall, Harvey, Heim, Kafka, Kimura, Larsen, Lee, Maryboy, Neilson, Nesvold,
  Simons, Sweitzer, Umpierre and Walker}]{SATCON2CEWG}
Venkatesan A, Lowenthal J, Arion D, Avila~Castro F, Bannister M, Barentine J,
  Begay D, Chavez JC, Carttar S, Gering R, Hartley R, Hall J, Harvey A, Heim J,
  Kafka S, Kimura K, Larsen K, Lee A, Maryboy N, Neilson H, Nesvold E, Simons
  D, Sweitzer J, Umpierre D and Walker C (2021{\natexlab{a}}) {SATCON2}
  {Community} {Engagement} {Working} {Group}.
\newblock Technical Report 10.5281/zenodo.5608920, American Astronomical
  Society.
\newblock \doi{10.5281/ZENODO.5608920}.
\newblock \urlprefix\url{https://zenodo.org/record/5608920}.

\bibitem[{Venkatesan et~al.(2021{\natexlab{b}})Venkatesan, Lowenthal, Arion,
  Avila~Castro, Bannister, Barentine, Begay, Chavez, Carttar, Gering, Hartley,
  Hall, Harvey, Heim, Kafka, Kimura, Larsen, Lee, Maryboy, Neilson, Nesvold,
  Simons, Sweitzer, Umpierre and Walker}]{Venkatesan2021}
Venkatesan A, Lowenthal J, Arion D, Avila~Castro F, Bannister M, Barentine J,
  Begay D, Chavez JC, Carttar S, Gering R, Hartley R, Hall J, Harvey A, Heim J,
  Kafka S, Kimura K, Larsen K, Lee A, Maryboy N, Neilson H, Nesvold E, Simons
  D, Sweitzer J, Umpierre D and Walker C (2021{\natexlab{b}}) {SATCON2}
  {Community} {Engagement} {Working} {Group}.
\newblock Technical Report 10.5281/zenodo.5608920, Zenodo.
\newblock \doi{10.5281/ZENODO.5608920}.
\newblock \urlprefix\url{https://zenodo.org/record/5608920}.

\bibitem[{Walker and Benvenuti(2022)}]{DQSIIWGR}
Walker C and Benvenuti P (2022) Dark and {Quiet} {Skies} {II} {Working} {Group}
  {Reports}.
\newblock Technical document 10.5281/zenodo.5874725, International Astronomical
  Union.
\newblock \doi{10.5281/ZENODO.5874725}.
\newblock \urlprefix\url{https://zenodo.org/record/5874725}.

\bibitem[{Walker and Hall(2020{\natexlab{a}})}]{SATCON1APX}
Walker C and Hall J (2020{\natexlab{a}}) Appendices to ``{Impact} of
  {Satellite} {Constellations} on {Optical} {Astronomy} and {Recommendations}
  {Toward} {Mitigations}''.
\newblock Technical report, American Astronomical Society.
\newblock
  \urlprefix\url{https://aas.org/sites/default/files/2020-08/SATCON1-WG-Tech-Reports_0.pdf}.

\bibitem[{Walker and Hall(2020{\natexlab{b}})}]{SATCON1}
Walker C and Hall J (2020{\natexlab{b}}) Impact of {Satellite} {Constellations}
  on {Optical} {Astronomy} and {Recommendations} {Toward} {Mitigation}
  ({SATCON1}).
\newblock Technical Report 10.3847/25c2cfeb.346793b8, American Astronomical
  Society.
\newblock
  \urlprefix\url{https://aas.org/sites/default/files/2020-08/SATCON1-Report.pdf}.

\bibitem[{Witze(2022)}]{Witze2022}
Witze A (2022) `unsustainable': how satellite swarms pose a rising threat to
  astronomy.
\newblock \emph{Nature} 606(7913): 236--237.
\newblock \doi{10.1038/d41586-022-01420-9}.
\newblock \urlprefix\url{https://doi.org/10.1038/d41586-022-01420-9}.

\bibitem[{Zhang(2020)}]{Zhang2020}
Zhang E (2020) {SpaceX}'s dark satellites are still too bright for astronomers
  (10 {September} 2020).
\newblock \emph{Scientific American}
  \urlprefix\url{https://www.scientificamerican.com/article/spacexs-dark-satellites-are-still-too-bright-for-astronomers/}.

\end{thebibliography}
 \bibliographystyle{SageH}
\small

\noindent 

\end{document}